\documentclass[11pt]{article}
\pdfoutput=1 
\usepackage{amssymb}
\usepackage{amsmath}
\usepackage{amstext}
\usepackage{graphicx,epsfig}
\usepackage{epsfig}
\usepackage{verbatim} 
\usepackage{fancybox}
\usepackage{color}
\usepackage{ulem}
\usepackage{enumitem}
\usepackage{youngtab}
\usepackage{subfigure}
\usepackage{bbm}
\usepackage{parskip}
\usepackage[numbers,sort&compress]{natbib}
\usepackage{ytableau}
\usepackage[utf8]{inputenc}

\usepackage{physics}
\usepackage{ulem}

\newcommand{\Comment}[1]{{}}
\definecolor{darkblue}{rgb}{0.15,0.35,0.55}
\definecolor{reddish}{rgb}{0.65, 0.2, 0.2}
\usepackage[linktocpage=true]{hyperref}
\hypersetup{
colorlinks=true,
citecolor=darkblue,
linkcolor=reddish,
urlcolor=darkblue,
pdfauthor={},
pdftitle={},
pdfsubject={}
}

\newcommand{\be}{\begin{equation}}
\newcommand{\ee}{\end{equation}}
\newcommand{\bea}{\begin{eqnarray}}
\newcommand{\eea}{\end{eqnarray}}
\newcommand{\beas}{\begin{eqnarray*}}
\newcommand{\eeas}{\end{eqnarray*}}

\definecolor{darkred}{rgb}{0.7,0.3,0.3}
\definecolor{darkgreen}{rgb}{0.2,0.7,0.3}
\definecolor{lightgreen}{rgb}{.816,.94,.753}
\definecolor{greyish}{rgb}{.8,.8,.8}
\definecolor{darkblue2}{rgb}{0.3,0.4,0.9}
\usepackage{pifont}
\usepackage{xcolor,colortbl}

\def\({\left(}
\def\){\right)}

\newcommand{\Mpl}{{{M_{\mathrm{Pl}}}}}
\newcommand{\Mplt}{{{M^2_{\mathrm{Pl}}}}}

\def\gsim{ \lower .75ex \hbox{$\sim$} \llap{\raise .27ex \hbox{$>$}} }
\def\lsim{ \lower .75ex \hbox{$\sim$} \llap{\raise .27ex \hbox{$<$}} }

\def\xyma{\xymatrix@M.7em}
\def\xymas{\xymatrix@M.1em}

\title{}
\author{}

\numberwithin{equation}{section}

\begin{document}

\renewcommand{\thefootnote}{\fnsymbol{footnote}}
~

\begin{center}
{\huge \bf Mapping the Weak Field Limit of Scalar-Gauss-Bonnet Gravity}
\end{center} 

\vspace{1truecm}
\thispagestyle{empty}
\centerline{\Large 
Benjamin Elder\footnote{\href{mailto:bcelder@hawaii.edu}{\texttt{bcelder@hawaii.edu}}}
and Jeremy Sakstein\footnote{\href{mailto:sakstein@hawaii.edu}{\texttt{sakstein@hawaii.edu}}}
}

\vspace{.5cm}
 
\centerline{{\it Department of Physics and Astronomy, University of Hawai'i,}}
 \centerline{{\it 2505 Correa Road, Honolulu, HI 96822, USA}} 

 \vspace{1cm}
\begin{abstract}
\noindent
We derive the weak field limit of scalar-Gauss-Bonnet theory and place novel bounds on the parameter space using terrestrial and space-based experiments. In order to analyze the theory in the context of a wide range of  
experiments, we compute the deviations from Einstein gravity around source masses with planar, cylindrical, and spherical symmetry. We find a correction to the Newtonian potential around spherical and cylindrical sources that can be larger than PPN corrections sufficiently close to the source. We use this to improve on laboratory constraints on the scalar-Gauss-Bonnet coupling parameter $\Lambda$ by two orders of magnitude. Present laboratory and Solar System bounds reported here are superseded by tests deriving from black holes.
\end{abstract}

\renewcommand*{\thefootnote}{\arabic{footnote}}
\setcounter{footnote}{0}

\section{Introduction}

Despite the many successes of general relativity (GR), we still do not know whether it is the full theory of gravity.  Although this theory is compatible with all direct tests to date, the persistence of several unexplained phenomena such as the cosmological constant problem~\cite{Padilla:2015aaa,Burgess:2013ara,Khoury:2018vdv} and the accelerated cosmological expansion~\cite{SupernovaSearchTeam:1998fmf}
has motivated the introduction of a wide range of alternative models to, and extensions of, GR~\cite{Clifton:2011jh, Joyce:2014kja,Baker:2019gxo}. Likewise, ongoing and ever-more precise tests of gravity in the laboratory, the Solar System, and in space are searching for hints of new physics beyond GR \cite{Adelberger:2002ic,Wagner:2012ui,Murphy:2012rea,Murphy:2013qya,Burrage:2016bwy,Burrage:2017qrf,Sakstein:2017pqi,CANTATA:2021ktz,Brax:2021wcv}.

Any extension of GR will increase the number of degrees of freedom in the theory beyond the nominal two \cite{Weinberg:1964ew,Weinberg:1965rz}. From this perspective, a minimal modification of gravity is to explicitly add a single degree of freedom in the form of a new scalar field $\phi$.  The scalar's couplings to the metric tensor, matter fields, and itself determine a wide range of possible phenomenologies \cite{Clifton:2011jh,Joyce:2014kja,Burrage:2017qrf,Baker:2019gxo}. Infra-red (IR) modifications of gravity  in which there is a direct coupling to the Ricci scalar $\phi R$ or, equivalently, a Yukawa coupling to matter fields $ \phi \bar \psi \psi$, result in \textit{fifth force} whose range is set by the inverse mass of the scalar field. This force modifies the Newtonian $1/r^2$ force law and there are ongoing searches for this modification from microscopic to cosmological scales~\cite{Adelberger:2003zx,Will:2005va,Koyama:2015vza}. Alternatively, ultra-violet (UV) modifications of gravity where the scalar-graviton couplings are non-renormalizable,
for example $\phi R^2, \phi R_{\mu \nu} R^{\mu \nu}$, give rise to deviations from GR in strongly gravitating systems e.g. black holes, neutron stars, and binary pulsars \cite{Berti:2015itd}.

In this work, we investigate the prospect for constraining UV modifications of GR focusing on scalar-Gauss-Bonnet gravity (SGB) in which a scalar field couples linearly to the Gauss-Bonnet topological invariant\footnote{We use units in which $c = \hbar = 1$ and have defined the reduced Planck mass as $\Mpl \equiv (8 \pi G)^{-1/2}$. }
\begin{equation}
    {\cal G} = R^2 - 4 R_{\mu \nu} R^{\mu \nu} + R_{\mu \nu \alpha \beta} R^{\mu \nu \alpha \beta}.
\end{equation}
Such theories are extremely interesting from a theoretical perspective for several reasons. First, couplings of the form $\phi \mathcal{G}$ represent the leading-order scalar-graviton interaction in shift-symmetric theories \cite{Yagi:2015oca}.  Second, the equations of motion resulting from such couplings are second-order, meaning that the theory does not suffer from an Ostrogradski ghost instability.~Third, more generalized couplings of the form $f(\phi)\mathcal{G}$ arise naturally in string theory \cite{Metsaev:1987zx,Kanti:1995vq}, may explain the accelerated expansion of the universe~\cite{Esposito-Farese:2003jfi, Odintsov:2021nim}, and give rise to the novel phenomenomenon of spontaneous black hole scalarization \cite{Silva:2017uqg,Kanti:1995vq,Antoniou:2017acq}. Studying the simplest coupling $\phi\mathcal{G}$ will lay the foundation for constraining these more complicated theories. 

The effects of SGB gravity are most pronounced in the strong-field regime, and hence black holes are powerful probes of this theory~\cite{Yunes:2011we,Yagi:2012gp, Perkins:2021mhb, Wang:2021yll}. This is helped by the fact that although smooth extended objects cannot obtain a scalar charge in the theory, black holes can~\cite{Yagi:2011xp,Yagi:2015oca}. Furthermore, certain types of couplings can be restricted based on theoretical arguments alone~\cite{Herrero-Valea:2021dry}.
It is natural to wonder how laboratory and Solar System experiments compare against astrophysical tests of this theory.  There currently exists a wide range of tabletop experiments that employ radically different source mass geometries to test gravity in the weak field regime.  Experiments like torsion balances~\cite{Lee:2020zjt}, atom interferometers~\cite{Biedermann:2014jya, Rosi:2014kva}, and Casimir force sensors~\cite{Chen:2014oda, Elder:2019yyp, Sedmik:2021iaw} have proven in recent years to be extraordinarily useful thanks to their high accuracy and their ability to be tuned to search for effects in specific theories (see \cite{Murata:2014nra} for a review of laboratory tests of gravity). For instance, large atom interferometers have performed some of the most precise measurements to date on Newton's constant $G$~\cite{Rosi:2014kva}, while miniature ones have proven sensitive to screened modified gravity theories that are otherwise very difficult to constrain~\cite{Hamilton:2015zga, Elder:2016yxm, Jaffe:2016fsh, Burrage:2016bwy, Sabulsky:2018jma}.

Some of the first experimental constraints on SGB gravity were derived from Solar System tests and focused on the specific case in which the scalar field drives the accelerated expansion of the universe~\cite{Esposito-Farese:2003jfi}. The PPN expansion of the theory has been computed, and it was found that the theory is indistinguishable from GR at second post-Newtonian order \cite{Sotiriou:2006pq}. This does not mean that the theory is impossible to constrain via local tests of gravity, only that the theory does not fit into the PPN framework.  Deviations from GR were computed around point particles in~\cite{Amendola:2007ni}, which were then used to place bounds from Solar System and laboratory tests. In this work we relax the point particle assumption, which enables us to use a larger range of experimental tests to constrain the theory. Specifically, we compute the weak field limit of SGB gravity and study the deviations from GR around extended objects with planar, cylindrical, and spherical symmetry. We find a $1/r^8$ force around spherical objects, a $1/r^5$ force around cylindrical objects, and no modification beyond GR around planar objects.  
We use results from a recent atom interferometry experiment to improve on the bound on the scalar-Gauss Bonnet coupling by two orders of magnitude relative to previous studies on laboratory and Solar System tests. Our bounds are still weaker than those deriving from black holes by 13 orders of magnitude. It is unlikely that even future experiments will be able to reach the same sensitivity as strong-field tests. 

The rest of this paper is organized as follows. In Section \ref{sec:SGB_theory} we briefly review the salient features of SGB gravity. In Section \ref{sec:weak-field} we expand the theory in the weak field limit, and compute its leading-order deviations from GR around extended bodies with planar, cylindrical, and spherical symmetry.  We also provide a simplified proof showing that extended bodies do not acquire a scalar charge in the theory, the details of which are in Appendix~\ref{app:scalar-charge}. In Section \ref{sec:experimental_bounds} we compare our bounds to those coming from black holes. We discuss the implications of our results and conclude in Section \ref{sec:conclusions}.

\section{Scalar-Gauss-Bonnet Gravity}
\label{sec:SGB_theory}

The action for SGB gravity that we will study is 
\begin{equation}
\label{eq:SGB_action}
    S = \int d^4x \sqrt{-g} \left( \frac{\Mplt}{2} R - \frac{1}{2} (\partial_\mu \phi)^2 +\frac{\phi}{\Lambda}\mathcal{G}+{\cal L}_\mathrm{SM}[g_{\mu\nu}; \psi]  \right),
\end{equation}
where $\Lambda$ is a new mass scale that parameterizes the SGB coupling, $\mathcal{L}_{\rm SM}$ represents the Standard Model Lagrangian, and the Standard Model matter fields $\psi$ couple minimally to the metric $g_{\mu\nu}$. Currently, the strongest bound on $\Lambda$ is $\Lambda > 1.2 \times 10^{-48}$ coming from black hole inspirals \cite{Perkins:2021mhb}. The action above describes the theory of a  shift-symmetric, parity-even\footnote{Taking the field to instead be a pseudo-scalar results in a coupling between the scalar and the Pontryagin density of the form $\phi \tilde{R}_{\mu\nu\alpha\beta}R^{\mu\nu\alpha\beta}$ where $\tilde{R}_{\mu\nu\alpha\beta}$ is the dual Riemann tensor. We will not study such couplings --- referred to as Chern-Simons couplings \cite{Alexander:2009tp} --- in this work.}  scalar coupled to gravity. The resulting equations of motion are second-order so there is no Ostragradski ghost instability and the theory propagates precisely three degrees of freedom (the two helicity-2 modes of the graviton and one helicity-0 scalar mode). Note that we have not included higher-dimensional scalar self-interactions or other shift-symmetric scalar-graviton couplings since these lead to high-order equations of motion and suffer from the Ostragradski instability. The status of SGB gravity as an effective field theory (EFT) is not well-studied and we will not attempt to do so in this work\footnote{Reference \cite{Yagi:2015oca} have found that SGB equations of motion are well-posed when the theory is treated as an EFT and expanded in the coupling parameter.}. One can break the shift symmetry, which then allows for the addition of a scalar potential or a generalized coupling of the form $f(\phi)\mathcal{G}$. We will not study these generalized couplings in this work but we briefly comment on them in Section \ref{sec:conclusions}.

\section{Weak field limit}
\label{sec:weak-field}

In this section we derive the non-relativistic, weak-field limit of SGB theory in the weak-coupling regime $\phi/\Lambda\ll1$. We then proceed to solve for the gravitational fields around highly symmetric source objects. We introduce a  bookkeeping parameter $\lambda$ into the action (that we will later set to unity) \eqref{eq:SGB_action} so that our action reads
\begin{align} 
    S=  \int d^4 x \sqrt{-g} \left(\frac{M_\mathrm{Pl}^2}{2} R - \frac{1}{2} (\partial_\mu \phi)^2 + \frac{\lambda}{\Lambda} \phi\,\! {\cal G} 
    + {\cal L}_\mathrm{SM}[g_{\mu\nu}; \psi] 
    \right)~.
\end{align}
We expand this action in the weak-field limit about flat spacetime, choosing the Newtonian gauge:
\begin{equation}
    ds^2 = - (1 + 2 \Phi) dt^2 + (1 - 2 \Psi) d \vec x^2~,
\end{equation}
where $|\Phi|, |\Psi| \ll 1$ are small, time-independent perturbations. In this limit we find that the Einstein-Hilbert $S_{\rm EH}$ and scalar $S_\phi$ (inluding the Gauss-Bonnet coupling) parts of the action are, respectively,
\begin{align} \nonumber
    S_\mathrm{EH} &= \int d^4 x \left( \Mpl^2 (\vec \nabla \Psi)^2 - 2 \Mpl^2 \vec \nabla \Psi \cdot \vec \nabla \Phi \right), \\
    S_\mathrm{\phi} &= \int d^4 x \left[ - \frac{1}{2} ( \vec \nabla \phi)^2 - \frac{1}{2} (\Phi - \Psi)(\vec \nabla \phi)^2 + 8 \lambda \frac{\phi}{\Lambda} \left( \nabla_i \nabla_j \Psi \nabla^i \nabla^j \Phi - \vec \nabla^2 \Psi \vec \nabla^2 \Phi \right) \right].
\end{align}
The expansion of the matter Lagrangian about flat space is:
\begin{align} 
    S_\mathrm{SM} = 
    \int d^4 x \left( {\cal L}_\mathrm{SM}[\psi] - \frac{\sqrt{-g}}{2} T_{\mu \nu} \left( g^{\mu \nu} - \eta^{\mu \nu} \right) \right).
    \label{grav-weakfield-action}
\end{align} 
If the matter is pressureless and non-relativistic then $T_{00} = \rho$ is the only non-zero component and we have
\begin{equation}
    S_\mathrm{SM} = \int d^4 x \left( {\cal L}_\mathrm{SM}[\psi] - \rho \Phi \right),
    \label{SM-weakfield-action}
\end{equation}
where the first term is the flat-space matter action and the second term its coupling to gravity. From the above expression it immediately follows that the action for a non-relativistic point particle $\rho = m \delta^3(\vec x)$ is
\begin{equation}
    S_\mathrm{pp} = \int d t \left( \frac{1}{2} m \dot {\vec x}^2 - m \Phi \right)~.
\end{equation}
The equation of motion for the point particle is the familiar 
\begin{equation}
    \ddot x_\mathrm{pp} = - \vec \nabla \Phi~.
    \label{pp-eom}
\end{equation}

Varying Eq.~\eqref{grav-weakfield-action} with respect to the fields $\phi$, $\Phi$, and $\Psi$ gives the gravitational field equations
\begin{align} \nonumber
    \vec \nabla^2 \Psi &= \frac{1}{2 \Mpl^2} \left( \rho + \frac{1}{2} (\vec \nabla \phi)^2 - 8 \frac{\lambda}{\Lambda} \left( \nabla^i \nabla^j (\phi \nabla_i \nabla_j \Psi) - \vec \nabla^2 (\phi \vec \nabla^2 \Psi) \right) \right),\\ \nonumber
    \vec \nabla^2 \Phi &= \vec \nabla^2 \Psi - \frac{1}{2 \Mpl^2} \left( \frac{1}{2} (\vec \nabla \phi)^2 + 8 \frac{\lambda}{\Lambda} \left( \nabla^i \nabla^j (\phi \nabla_i \nabla_j \Phi) - \vec \nabla^2 (\phi \vec \nabla^2 \Phi) \right) \right),\\
    \vec \nabla^2 \phi &= \vec \nabla \cdot \left((\Psi - \Phi) \vec \nabla \phi \right) - 8 \frac{\lambda}{\Lambda} \left( \nabla_i \nabla_j \Psi \nabla^i \nabla^j \Phi - \vec \nabla^2 \Psi \vec \nabla^2 \Phi \right).
\label{sGB-EOMS}
\end{align}
Our task is to solve for the gravitational fields around an extended body, and then use Eq.~\eqref{pp-eom} to compute the motion of a test particle.  Given the difficulty of Eq.~\eqref{sGB-EOMS}, it is necessary to solve this system of equations order-by-order in $\lambda$ i.e., by expanding
\begin{align} \nonumber
    \Phi &= \Phi_0 + \Phi_1 (\lambda) + \Phi_2 (\lambda^2) + {\cal O} (\lambda^3)~, \\
    \phi &= \phi_0 + \phi_1 (\lambda) + {\cal O} (\lambda^2) + ...
\end{align}
where a subscript $n$ indicates the order of the expansion in  $\lambda^n$. We emphasize that this is {\it not} a post-Newtonian expansion, as we have assumed the metric is linear in the gravitational potentials.  Rather, this is an expansion of the Gauss-Bonnet terms in the weak-coupling limit.  We will later confirm that dropping the PPN terms beyond leading order is consistent.

At leading order in the Gauss-Bonnet expansion we have $\lambda = 0$, and the system of equations is
\begin{align} \nonumber
\vec \nabla^2 \Psi_0 &= 4 \pi G \left( \rho + \frac{1}{2} (\vec \nabla \phi_0)^2 \right), \\ \nonumber
\vec \nabla^2 \Phi_0 &= \vec \nabla^2 \Psi_0 - 4 \pi G \left( \frac{1}{2} (\vec \nabla \phi_0)^2 \right), \\
\vec \nabla^2 \phi_0 &= \vec \nabla \cdot \left( (\Psi_0 - \Phi_0) \vec \nabla \phi_0 \right).
\end{align}
In the ensuing subsections we will examine this system of equations in situations where the mass distribution exhibits planar, cylindrical, and spherical symmetry.

\subsection{Planar symmetry}
\label{subsec:planar}
The planar configuration is trivial.  In this case, all fields depend only on a single spatial coordinate, so the terms proportional to $\lambda$ in Eq.~\ref{sGB-EOMS} all vanish.  The scalar field is therefore decoupled from gravity and we are left with ordinary Newtonian gravity, at least at the lowest orders in $\lambda$.  The cylindrical and spherical cases require more care and are treated in the next subsections.

\subsection{Spherical symmetry}

We now specialize to a spherically symmetric body i.e., $\rho = \rho(r)$.  At lowest order the gravitational fields are sourced by the ordinary matter distribution, so we have $\Phi = \Psi$.  At lowest order we also take $\lambda = 0$ so the third equation in Eq.~\eqref{sGB-EOMS} is solved by setting $\phi_0$ to be a constant that we can set to zero using the shift symmetry, leaving us with the usual Newtonian system
\begin{equation}
    \vec \nabla^2 \Psi_0 = \vec \nabla^2 \Phi_0 = 4 \pi G \rho(r).
\end{equation}
This can be integrated to give
\begin{equation}
    \Phi_0' = \Psi_0' = \frac{G m(r)}{r^2}~,
\end{equation}
where we have defined the enclosed mass
\begin{equation}
    m(r) \equiv \int_0^r 4 \pi r' \rho(r') dr'.
\end{equation}
There is no need to integrate the equations once more to find $\Phi_0$ and $\Psi_0$ as the equations of motion depend only on first and second derivatives of $\Phi_0, \Psi_0$.

Using the solutions for $\phi_0$, $\Psi_0$, and $\Phi_0$ above, we can derive the equation of motion  for $\phi_1$. We find
\begin{equation}
    \vec \nabla^2 \phi_1 = - \frac{8 \lambda}{\Lambda} \left( \nabla_i \nabla_j \Psi_0 \nabla^i \nabla^j \Phi_0 - \vec \nabla^2 \Psi_0 \vec \nabla^2 \Phi_0 \right).
    \label{phi1-eom-general}
\end{equation}
Making use of the following identity for two functions of radius $f(r), g(r)$ in spherical coordinates,
\begin{equation}
    \nabla_i \nabla_j f(r) \nabla^i \nabla^j g(r) = f'' g'' + \frac{2}{r^2} f' g',
    \label{cross-deriv-spherical}
\end{equation}
the equation of motion simplifies to
\begin{equation}
    \partial_r (r^2 \phi'_1) = \frac{16 \lambda G^2}{\Lambda} \partial_r \left( \frac{m^2}{r^3} \right).
\end{equation}
This may be integrated once to give
\begin{equation}
    \phi_1'(r) = \frac{16 \lambda }{\Lambda} \frac{G^2 m^2}{r^5} + \frac{C}{r^2},
    \label{phi1-sphere}
\end{equation}
with integration constant $C$. We demand that $\phi$ is regular at the origin i.e., $\phi'(0) = 0$.  In the small-$r$ limit we have $m \sim r^3$~, so the first term vanishes automatically, and we are left with the constraint $C = 0$.
There are no corrections to the gravitational potentials $\Phi, \Psi$ at $\mathcal{O}(\lambda)$: $\Phi_1 = \Psi_1 = 0$.  

The leading SGB correction to the metric potentials arises at $\mathcal{O}(\lambda^2)$. 
The equation of motion for $\Phi_2$ is obtained by adding the first two equations of Eq.~\eqref{sGB-EOMS} together and expanding the derivatives, yielding
\begin{equation}
    \vec \nabla^2 \Phi_2 = -\frac{ 8}{\Mpl^2} \frac{\lambda}{\Lambda} \left( \nabla^i \nabla^j \phi_1 \nabla_i \nabla_j \Phi_0 - \vec \nabla^2 \phi_1 \vec \nabla^2 \Phi_0 \right).
    \label{Phi2-eom-general}
\end{equation}
Using Eq.~\eqref{cross-deriv-spherical} and our solutions for $\Phi_0, \phi_1$, this simplifies to
\begin{equation}
    \vec \nabla^2 \Phi_2 =  \frac{  2^{11}\pi \lambda^2 G^4}{\Lambda^2} \frac{1}{r^2} \partial_r \left( \frac{m^3}{r^6} \right)~.
\end{equation}
Integrating once, we obtain
\begin{equation}
    \Phi_2' = \frac{ 2^{11} \pi \lambda^2 G^4 }{\Lambda^2} \frac{m^3}{r^8}~.
    \label{Phi2-sph}
\end{equation}
Once again the monopole term proportional to $1/r$ has vanished by demanding regularity at the origin, just as in Eq.~\eqref{phi1-sphere}.  This potential gradient, via the geodesic equation Eq.~\eqref{pp-eom}, gives the leading scalar-Gauss-Bonnet acceleration of a test particle in the vicinity of a spherical source mass.  This result was obtained previously in \cite{Amendola:2007ni} via a slightly different route.

Recall that we are working in the limit in which the sGB contribution is larger than the PPN contributions.  In spherical coordinates, the leading-order post-Newtonian correction is
\begin{equation}
    \Phi_\mathrm{1PN}' = \frac{G^2 m^2}{r^3}.
\end{equation}
This decays with $r$ more slowly than the leading scalar-Gauss-Bonnet contribution given by Eq.~\eqref{Phi2-sph}. Nevertheless, the scalar-Gauss-Bonnet coupling may dominate on macroscopic scales provided that $\Lambda$ is sufficiently small. Specifically, the SGB potential dominates over the 1PN potential so long as
\begin{align} 
    \Lambda &\ll \left( \frac{2^{11} \pi G^2 m}{r^5} \right)^{1/2}. 
    \label{Lambda-validity}
\end{align}
This limit will need to be verified before placing bounds from a given experiment.  This will be discussed in more detail in Section~\ref{sec:experimental_bounds}.

Before considering non-spherical sources, we briefly pause to remark that, as expected, that our spherical source does not carry a scalar monopole charge (meaning there is no term that scales as $Q/r$ as $r\rightarrow\infty$.). Although new scalar fields often do imbue objects with scalar charge, fields coupled only to the Gauss-Bonnet density do not. This has been proven for extended objects~\cite{Yagi:2015oca}, with black holes as a notable exception. We present a simplified version of that proof, specialized to spherically symmetric systems, in Appendix~\ref{app:scalar-charge}.

\subsection{Cylindrical geometry}
We now repeat the analysis for the case in which the matter distribution has cylindrical symmetry $\rho = \rho(r)$, with $r$ the radial distance in cylindrical coordinates.  We begin by defining the mass per unit length
\begin{equation}
    m(r) = \int_0^r 2 \pi r \rho(r) dr.
\end{equation}
Then, at leading order $\lambda = 0$ and we have
\begin{align} \nonumber
    \Phi_0' &= \Psi_0' = \frac{2 G m}{r} + \frac{C}{r}, \\
    \phi_0 &= 0,
\end{align}
with $C$ an integration constant.  As before, demanding regularity at the origin imposes $C = 0$.  In cylindrical coordinates we have the identity
\begin{equation}
    \nabla_i \nabla_j f(r) \nabla^i \nabla^j g(r) = f'' g'' + \frac{1}{r^2} f' g',
    \label{cross-deriv-cyl}
\end{equation}
which allows us to simplify Eq.~\eqref{phi1-eom-general} to
\begin{equation}
    \vec \nabla^2 \phi_1 = \frac{8 \lambda}{\Lambda} \frac{2}{r} \Phi_0'' \Phi_0'.
\end{equation}
Using the solutions for $\Phi_0$ we can bring this into the form
\begin{equation}
    \vec \nabla^2 \phi_1 = \frac{1}{r} \frac{32 \lambda G^2}{\Lambda} \partial_r \left( \frac{m^2}{r^2} \right),
\end{equation}
which can readily be integrated to yield
\begin{equation}
    \phi_1' = \frac{32 \lambda G^2}{\Lambda} \frac{m^2}{r^3}~,
\end{equation}
where, as usual, we have discarded the monopole term by demanding regularity of $\phi_1$ at the origin.

Finally we turn to the solution of $\Phi_2$.  Using Eqs.~\eqref{Phi2-eom-general} and~\eqref{cross-deriv-cyl}, we have
\begin{equation}
    \vec \nabla^2 \Phi_2 = \frac{8 \lambda}{\Mpl^2 \Lambda} \frac{1}{r} \partial_r \left( \phi_1' \Phi_0' \right),
\end{equation}
which integrates once to
\begin{equation}
    \Phi_2' = \frac{2^{12} \pi G^4 \lambda^2}{\Lambda^2} \frac{m^3}{r^5}.
    \label{cylinder-force}
\end{equation}
This gives the motion of a test particle due to the scalar-Gauss-Bonnet coupling in cylindrically symmetric geometries.

Once again we must compare this to the leading Post-Newtonian corrections.  In cylindrical coordinates, this is
\begin{equation}
    \Phi_\mathrm{1PN} = 4 G^2 m^2 \ln(r)^2~.
\end{equation}
Differentiating, and comparing to Eq.~\eqref{cylinder-force}, the sGB contribution is larger provided that
\begin{equation}
    \Lambda^2 < \frac{2^9 \pi G^2 m}{r^4 \ln r}~.
\end{equation}

\section{Experimental Bounds}
\label{sec:experimental_bounds}
In this section we derive the bounds that may be placed on the theory from various local experimental tests using the results for the scalar field profiles found in the previous section. These results are summarized in Table~\ref{tab:summary}, and we henceforth set the bookkeeping parameter $\lambda=1$. Experimental bounds on new physics are often framed in terms of either a new Yukawa-type interaction or in the PPN expansion.  The scalar-Gauss-Bonnet force Eq.~\eqref{Phi2-sph} does not fit into either of these frameworks, so we must carefully reinterpret individual constraints.  Tests outside the Solar System have been considered elsewhere and are briefly discussed in Appendix~\ref{app:strong-gravity}.

\renewcommand{\arraystretch}{2}
\begin{table}[t]
\centering

\begin{tabular}{| c || c | c | c | c} \hline
    geometry    & planar & cylindrical & spherical \\ \hline 
      $a_5$ & 0 & $\frac{2^{12} \pi G^4 }{\Lambda^2} \frac{m^3}{r^5}$ & $\frac{ 2^{11} \pi  G^4 }{\Lambda^2} \frac{m^3}{r^8}$ \\
      \hline
\end{tabular}
\caption{\small Summary of the leading scalar-Gauss-Bonnet corrections to the motion of a test particle around a massive source in the weak-field limit. $a_5$ is the SGB fifth force (per unit mass) and we remind the reader that with spherical symmetry $m$ is the total mass of the source, while in cylindrical symmetry $m$ is the mass per unit length.}
\label{tab:summary}
\end{table}

Finally, we note that experiments that use a planar (or mostly-planar) source mass are not sensitive to scalar-Gauss-Bonnet gravity.  As shown in the previous section, planar source masses do not source the scalar field and hence their gravitational field is identical to what it would be in Einstein gravity.  Experiments in this category include bouncing neutrons~\cite{Cronenberg:2018qxf}, Casimir sensors~\cite{Elder:2019yyp, Bimonte:2021sib}, and the E\"{o}t-Wash torsion balance experiment~\cite{Lee:2020zjt}, although torsion balances with non-planar geometries can still be considered~\cite{Amendola:2007ni}.

\subsection{Atom interferometry}
An atom interferometer measures the acceleration of an individual atom in free-fall.  Typically, an experiment will perform two measurements, with a source mass in a ``near'' and ``far'' configuration, allowing the force between the atom test mass and the source mass to be isolated.

A recent atom interferometry experiment \cite{Biedermann:2014jya} measured the acceleration of atoms towards a 1080 kg rectangular box-shaped source mass, with the atoms 20 cm from the surface.  The experiment had a resolution of $20 \times 10^{-11}~\mathrm{m/s}^2$, which was sufficient to clearly resolve the Newtonian gravitational force ($\sim 10^{-7}~\mathrm{m/s}^2$).  Approximating the box-shaped source mass as a sphere and using Eq.~\eqref{Phi2-sph}, we find that the experiment gives the constraint
\begin{equation}
    \Lambda > 2.9 \times 10^{-61}~\mathrm{eV}.
\end{equation}
Using Eq.~\eqref{Lambda-validity} we find that the validity of our calculation requires $\Lambda \ll 10^{-50}~\mathrm{eV}$, which is easily satisfied for this bound.

The authors of~\cite{Biedermann:2014jya} note that there are several upgrades to the experiment that are feasible, such as bringing the atoms closer to the source mass, increasing the averaging time to one month, and by increasing the vertical size of the source mass for a longer drop time of the atoms would increase the sensitivity by a factor of $\sim 600$.  With those improvements, such an experiment would be able to rule out an additional two orders of magnitude for a constraint of $\Lambda > 10^{-59}~\mathrm{eV}$.  For an experiment with a larger drop time, it is likely advantageous to employ a cylindrically symmetric source mass.  The interaction could therefore be estimated using Eq.~\eqref{cylinder-force}.  For the same experimental parameters, a tungsten cylinder with a 42~cm radius provides the same sensitivity, although this configuration allows for the possibility of much longer measurement times.

There are other notable measurements of gravity with atom interferometers. For example, the experiment described in \cite{Rosi:2014kva} is competitive with other atom interferometer measurements of Newton's constant, but  used a more complicated source mass geometry.  Producing a prediction of scalar-Gauss-Bonnet gravity in such a setup is beyond the scope of this work.

One could also look to miniaturized atom interferometers, which have recently been used in searches for modified gravity at the dark energy scale \cite{Jaffe:2016fsh,Sabulsky:2018jma}.  At first glance one might think that such experiments might be useful in searching for the short-ranged force of Eq.~\eqref{Phi2-sph}, as they use small source masses and thus probe short distance scales.  However, one finds that these experiments are limited in this case by the small size of the source masses. For a spherical source mass of with a given (constant) density $\rho$ and radius $R$, the scalar-Gauss-Bonnet force at the surface scales as $\rho^3 R$ (see Table~\ref{tab:summary}) hence it is advantageous to consider experiments employing source masses that are as large as possible.

\subsection{Atomic hydrogen spectroscopy}
In \cite{Brax:2014vva} it was shown that a disformally-coupled scalar field is constrained by measurements of the 1s~$\to$~2s transition in hydrogen.  The model considered there resulted in a force law that also scaled as $r^{-8}$, allowing us to interpret their bound into one on our coupling parameter $\Lambda$. The only required step is to write our variable $\Lambda$ in terms of their variables.

Based on Eq.~\eqref{Phi2-sph}, we see that in scalar-Gauss-Bonnet gravity the Coulomb potential is modified to become
\begin{equation}
    V(r) = - \frac{e^2}{r} - \frac{2^{11} \pi G^4}{7 \Lambda^2} \frac{m_\mathrm{e} m_\mathrm{p}^3}{r^7}~.
\end{equation}
Comparing this to Eq.~(6.2) in \cite{Brax:2014vva}, we find that 
\begin{equation}
    \Lambda = \frac{2^8 \pi^2 G^2 m_\mathrm{p}}{\sqrt{21}} M^4,
\end{equation}
where $m_\mathrm{e}, m_\mathrm{p}$ are the electron and proton masses, and the interaction in \cite{Brax:2014vva} was $T^{\mu\nu}\partial_\mu\partial_\nu\phi/M^4$. The hydrogen 1s $\to$ 2s transition yielded the constraint $M > 0.2~\mathrm{GeV}$ \cite{Schwob:1999zz, Jaeckel:2010xx}.  Using the above relation, we find that this translates to a constraint
\begin{equation}
    \Lambda > 4.2 \times 10^{-68}~\mathrm{eV}~.
\end{equation}
Computing the regime of validity with Eq.~\eqref{Lambda-validity} for a proton mass and a distance of approximately 1 Angstrom, we find that the PPN corrections are smaller than the sGB contribution so long as $\Lambda < 10^{-42}~\mathrm{eV}$.  As such, this constraint is well within the regime of validity of our analysis.

\subsection{Torsion Balance}
Torsion balances are a leading method to test the gravitational inverse square law and to search for new interactions.  One of the most precise torsion balances to date is the Eot-Wash experiment, although this turns out be unsuitably for our present analysis.  This is because that experiment uses a geometry that is far more complex than the symmetric configurations investigated in this paper.  Furthermore, that geometry is approximately planar, so it is expected that the scalar-Gauss-Bonnet force would be highly suppressed by this setup, as discussed in Sec.~\ref{subsec:planar}.

The best bounds on the scalar-Gauss-Bonnet theory coming from torsion balances originally appeared in~\cite{Amendola:2007ni} and was based on an experiment that measured the torque on a copper bar due to nearby spherical source masses~\cite{PhysRevD.32.3084}.  This yielded a bound, when translated into our variable $\Lambda$, of
\begin{equation}
    \Lambda > 2.5 \times 10^{-63}~\mathrm{eV}~.
\end{equation}

\subsection{Lunar Laser Ranging}
Measurements of the Earth-Moon distance, integrated over decades, has tested the $1/r^2$ force law on the moon to an accuracy of \cite{Murphy:2013qya}
\begin{equation}
    \frac{\delta a}{a_\mathrm{N}} < 2 \times 10^{-11}.
\end{equation}
Using Eq.~\eqref{Phi2-sph} to compute the Moon's anomalous acceleration $\delta a$~, and using an average Earth-Moon distance of $385,000~\mathrm{km}$ \cite{Murphy:2013qya},  we find a constraint
\begin{equation}
    \Lambda > 4.8 \times 10^{-63}~\mathrm{eV}~.
\end{equation}
Using Eq.~\eqref{Lambda-validity}, we find that this is right at the edge of the regime of validity for our calculation, which requires $\Lambda < 6.5\times 10^{-63}~\mathrm{eV}$. This bound should therefore be interpreted with caution, as a rigorous treatment would require the inclusion of post-Newtonian gravity. However, this bound is weaker than the one deriving from atom interferometry, so for our present purposes such an analysis will not be necessary. 

\subsection{Cassini}
In \cite{Amendola:2007ni} the gravitational time delay was computed for the Cassini spacecraft in scalar-Gauss-Bonnet gravity, finding $\sqrt{|\alpha|} < 8.9 \times 10^{11}~\mathrm{cm}$. Translating to our variables (see Eq.~\eqref{alpha-to-lambda} and its derivation in the next section) we obtain the constraint
\begin{equation}
    \Lambda > 4.2 \times 10^{-62}~\mathrm{eV}~.
\end{equation}
Apart from atom interferometry this is the strongest bound deriving from tests of gravity within the Solar System. 

\section{Discussion and Conclusions}
\label{sec:conclusions}

In this paper we have derived the weak-field limit of scalar-Gauss-Bonnet theory of gravity. This enabled us to compute the leading corrections to general relativity in a range of laboratory and Solar System tests. We find that the most constraining local test of the theory comes from atom interferometry, for which we find $\Lambda>2.9\times10^{-61}$ eV where the scalar-Gauss-Bonnet coupling is $\phi\mathcal{G}/\Lambda$.  Our constraint improves on existing laboratory bounds by two orders of magnitude.  These bounds, however, are weaker than those originating from tests of black holes by  thirteen orders of magnitude.  It is unlikely that terrestrial and Solar System tests will be competitive with strong-field probes in the near future, however we note that there may be some important exceptions to this. For example, some phenomena such as spontaneous black hole scalarization due to a quadratic scalar-Gauss-Bonnet coupling of the form $\phi^2\mathcal{G}/\Lambda^2$ require a mass and quartic self-interaction for the scalar in order to be stable \cite{Macedo:2019sem,Silva:2018qhn}. Such couplings dramatically reduce the range of the fifth force and it may be the case that laboratory experiments are the only probes sensitive to these short ranges (depending on the size of the mass and quartic coupling). A study of such couplings would be an interesting topic for future study.


{\bf Acknowledgments:}
We are grateful to Clare Burrage and Justin Khoury for helpful discussions. Some of the expressions in this work were derived with the aid of the Mathematica  package xAct~\cite{martin-garcia, Mart_n_Garc_a_2008, Brizuela:2008ra, Pitrou:2013hga}.

{\bf Software:}
Mathematica 12, xAct 1.1.4.


\appendix

\section{Scalar charge}
\label{app:scalar-charge}
In this appendix we show that extended, spherically symmetric bodies do not source a monopole scalar field term $\phi \sim \mu / r$, i.e. a scalar charge.  This argument is based on a more general proof given in~\cite{Yagi:2015oca}. We begin by considering the equation of motion of a scalar coupled only to the Gauss-Bonnet invariant in the action:
\begin{equation}
    \Box \phi = \frac{c}{\Lambda} {\cal G}~,
    \label{schematic-eom}
\end{equation}
where $c$ contains prefactors that are irrelevant to the present discussion.

We imagine that there is some localised, static source at the origin. This implies that the metric components decay as $1/r$ towards spatial infinity, so the components of the Riemann tensor scale at least as $1/r^2$.  We will also make use of the fact that, when the space-time is endowed with a Killing vector, the Gauss-Bonnet density may be written as a total derivative~\cite{Yale:2011usf}
\begin{equation}
    {\cal G} = \vec \nabla \cdot \vec J~,
\end{equation}
for some $\vec J$.  The precise form of $\vec J$ will not be important, but we do need to know its scaling with $r$.  The left hand side scales as $1/r^4$, implying that $|\vec J| \sim 1/r^3$.

Let us now assume that the source does have some scalar charge $\mu$, which means that the scalar field profile includes a $1/r$ term
\begin{equation}
    \phi = \frac{\mu}{r} + O\left(\frac{1}{r^2} \right)~.
\end{equation}

We now compute the volume integral of Eq.~\eqref{schematic-eom}.  We do this far from the source, so that we can use the flat space measure:
\begin{equation}
    \int_V \vec \nabla^2 \phi dV = \frac{c}{\Lambda} \int_V {\cal G} dV~,
\end{equation}
where $V$ is a spherical volume centered on the origin.  Considering the left-hand side first, we can use the divergence theorem to turn the volume integral into a surface integral:
\begin{equation}
    \mathrm{LHS} = \int_V \vec \nabla^2 \phi dV = \int_S \vec \nabla \phi \cdot \vec {d S}~.
\end{equation}
If the boundary is taken to be sufficiently far from the source, the gradient of the scalar field $\vec \nabla \phi$ is approximately normal to the spherical surface and we have
\begin{equation}
    \mathrm{LHS} = 4 \pi r^2 |\vec \nabla \phi| = 4 \pi \mu~.
\end{equation}
Considering the right-hand side, we rewrite $\cal G$ in terms of a total derivative:
\begin{equation}
    \mathrm{RHS} = \frac{c}{\Lambda} \int_V \vec \nabla \cdot \vec J dV = \int_S \vec J \cdot \vec{d S}~. 
\end{equation}
Once again, sufficiently far from the source $\vec J$ is approximately normal to the spherical surface, giving
\begin{equation}
    \mathrm{RHS} = \frac{c}{\Lambda} 4 \pi r^2 |\vec J| \sim \frac{1}{r}~.
\end{equation}
Matching left and right hand sides, and taking the limit $r \to \infty$ gives us the result
\begin{equation}
    \mu = 0~.
\end{equation}
Thus we find that a scalar field coupled only to the Gauss-Bonnet density does not imbue static, localised sources with scalar charge.  Black holes, however, are a notable exception to this argument~\cite{Yagi:2015oca}, as they are not extended bodies.

\section{Comparison with astrophysical strong gravity tests}
\label{app:strong-gravity}

Tests involving black holes are significant as, unlike extended objects, they can develop a scalar charge in scalar-Gauss-Bonnet theory.  This enables much stronger interactions between black holes in this theory.  Astrophysical studies of this model are typically written in units in which $c = G = 1$:
\begin{equation}
    S = \int d^4 x \sqrt{-g} \left(\frac{1}{16 \pi} R + \alpha \phi {\cal G} - \frac{1}{2} (\partial \phi)^2  + {\cal L}_\mathrm{mat}~ \right)~.
\end{equation}
These studies report constraints on the coupling $\alpha$, which is analogous to our $\Lambda$.  In these units, $\alpha$ has dimension of length$^2$ and $\phi$ is dimensionless.  In order to compare those constraints to the ones obtained here, it is necessary to convert to the particle physics units we use in this paper.

Moving now to our own units where $c = \hbar = 1$ and $G = (8 \pi \Mpl^2)^{-1}$, we begin by re-inserting factors of $G$ to match dimensions across the different terms:
\begin{equation}
    S = \int d^4 x \sqrt{-g} \left(\frac{1}{16 \pi G} R + \frac{1}{G} \alpha \phi {\cal G} - \frac{1}{G}\frac{1}{2} (\partial \phi)^2  + {\cal L}_\mathrm{mat}~ \right)~.
\end{equation}
Replacing $G$ with $\Mpl$ and canonically normalizing $\phi$, we find
\begin{equation}
    S = \int d^4 x \sqrt{-g} \left(\frac{\Mpl^2}{2} R + \sqrt{8 \pi} \Mpl \alpha \phi {\cal G} -\frac{1}{2} (\partial \phi)^2  + {\cal L}_\mathrm{mat}~ \right)~.
\end{equation}
This gives the conversion between different conventions for the coupling parameter:
\begin{equation}
    \Lambda = \frac{1}{\sqrt{8 \pi} \Mpl \alpha}~,
    \label{alpha-to-lambda}
\end{equation}
enabling us to convert a stated bound on $\alpha$ to our variable $\Lambda$.  The strongest bound to date derives from black hole inspirals~\cite{Perkins:2021mhb}, and places the constraint $\sqrt{|\alpha|} < 3.1 \times 10^{5} ~\mathrm{cm}$.  In terms of our units, this translates to 
\begin{equation}
    \Lambda > 3.5 \times 10^{-49}~\mathrm{eV}~,
\end{equation}
which is roughly 12 orders of magnitude stronger than the strongest local test of gravity.  It is notable that the mere existence of solar-mass sized black holes places a bound that is nearly as strong as this one~\cite{Yagi:2012gp}.

\renewcommand{\em}{}
\bibliographystyle{utphys}
\addcontentsline{toc}{section}{References}
\bibliography{main}


\end{document}